\documentstyle[aps]{revtex} 
\newcommand{\be}{\begin{equation}}
\newcommand{\ee}{\end{equation}}
\newcommand{\bea}{\begin{eqnarray}}
\newcommand{\eea}{\end{eqnarray}}
\begin{document}
\preprint{SNUTP 96-109, cond-mat/9610169}
\draft
\title{Asymmetric XXZ chain at the 
antiferromagnetic transition: Spectra and partition functions}

\author{ Doochul Kim\cite{email}}
\address{Department of Physics and Center for Theoretical Physics,
Seoul National University, Seoul 151-742, Korea} 
\date{October 23, 1996}
\maketitle
\begin{abstract}
The Bethe ansatz equation is solved to obtain analytically
the leading finite-size correction of the spectra of the 
 asymmetric XXZ chain and the accompanying isotropic 6-vertex model
 near the antiferromagnetic phase boundary 
at zero vertical field. The energy gaps scale with size $N$ as
$N^{-1/2}$ and their amplitudes are obtained 
in terms of level-dependent scaling functions. Exactly on the
phase boundary, the amplitudes are proportional to a sum of square-root 
of integers and an anomaly term. 
By summing over all low-lying levels, the partition functions are 
obtained explicitly. 
Similar analysis is performed also at the 
phase boundary of zero horizontal field in which case the energy
gaps scale as $N^{-2}$. The partition
functions for this case are
found to be that of a non-relativistic free fermion system.
 From symmetry of the lattice model under $\pi /2$ rotation,
several identities between the partition functions are found. 
The $N^{-1/2}$ scaling at zero vertical field is interpreted as
a feature arising from viewing the Pokrovsky-Talapov transition
with the space and  time coordinates interchanged. 

\end{abstract}


\section{Introduction}
\label{sec:intro}

Spectral properties of the asymmetric XXZ chain and its associated lattice
model, the asymmetric 6-vertex model, have been of much interest recently 
\cite{albdw,nk3,gwa,dennijs,kim95}. The asymmetric XXZ chain is  a
non-hermitian generalization of the standard spin 1/2 XXZ chain 
\cite{baxter} and is the anisotropic limit of the row-to-row transfer
matrix of the asymmetric 6-vertex model, i.e. the general 6-vertex model in
horizontal and vertical fields \cite{liebwu,nolden}. 
Both models are solvable
by the Bethe ansatz for arbitrary interaction parameter $\Delta$,
horizontal field $H$ and vertical field $V$, and for the lattice model,
also for arbitrary anisotropy parameter.

When the interaction parameter $\Delta$ is less that $-1$,
the asymmetric XXZ chain and the asymmetric 6-vertex model
are antiferromagnetically ordered for small
values of $H$ and $V$. (We use the term antiferromagnetic phase
to denote the antiferroelectric phase of the lattice model.)
 The ordered phase in the $H$-$V$ plane is bounded
by the antiferromagnetic phase boundary beyond which the systems are
disordered. The disordered phase is critical in that excitations are
massless
and correlation lengths decay algebraically. For still larger values
of fields, complete ferromagnetic ordering sets in. 
The free energy of the 6-vertex model as a function of the two fields can be
interpreted as the equilibrium crystal shape \cite{ecs,buks} and the
antiferromagnetically ordered region of fields corresponds to
 the flat facet,
and the antiferromagnetic phase boundary to the facet boundary.
On the facet boundary, 
the curvature of the free energy surface is universal 
\cite{akutsu,nk3}. The
nature of phase transitions at the facet boundary can be understood from
the view point of domain wall excitations and is that of the
Pokrovsky-Talapov (PT) transition \cite{pt,dennijs88}.

Within the critical phase and on the phase boundaries, there exist 
infinite number of massless excitations in the thermodynamic limit. 
When the system size, i.e. the number of spins in the chain $N$, is finite, 
the degeneracies are lifted producing energy gaps. Finite-size scaling of
these energy gaps gives valuable information on the properties of the
system \cite{fss}.  Throughout  the critical phase of the asymmetric
XXZ chain and the asymmetric 6-vertex model, 
 the energy gaps scale as $N^{-1}$, and as shown in \cite{nk3}, the
finite-size scaling amplitudes can be accounted for by the central charge
$c$=1 conformal field theory \cite{c=1}, with suitable modifications to
account for incommensurate arrow densities. All finite-size scaling 
information are encoded in the $O(1)$ part of the partition
function which is covariant under the modular transformations
and, for the case of the asymmetric 6-vertex model, is given
by the modified Coulombic partition function. 
 
When $\Delta > 1$ and $V=0$, the critical phase is bounded by the
so-called stochastic line where 
the asymmetric XXZ chain Hamiltonian describes time
evolution of the single-step model \cite{plischke} 
which is one of the simplest realization
of the Kardar-Parisi-Zhang (KPZ)
 universality class of non-equilibrium growth 
models \cite{kpz} and also of the driven lattice gas 
\cite{liggett,henkel,derrida}.
On the stochastic line, the $N^{-1}$ scaling crossovers to the KPZ
type $N^{-3/2}$ scaling where the exponent 3/2 is the dynamic
exponent for the 1+1 dimensional KPZ class.
This has been shown first in \cite{gwa} for
a special limit, and numerically in \cite{dennijs} for other cases.  
More recently, it was shown in \cite{kim95}, which will be denoted as I in
this work, that the energy gaps for general cases can be
systematically expanded in power series of $N^{-1/2}$ 
and the scaling amplitudes are given by level-dependent, but universal,
 scaling functions.

Recently, the finite-size scaling on the antiferromagnetic phase
boundary of $V=0$ is discussed by Albertini {\it et al.}
 \cite{albdw}. In their work, it was shown
analytically that the
low-lying excitations  satisfy the energy-momentum relation $\varepsilon
\sim \sqrt{ik}$ in the thermodynamic limit which in turn indicates
the $N^{-1/2}$ scaling of energy gaps. That the energy gaps scale as
$N^{-1/2}$ was then confirmed 
by numerical solutions of the Bethe ansatz equations for finite $N$. 

In this work, we present
analytic solutions for the finite-size scaling at the
antiferromagnetic phase boundary using the method developed in I. 
It is seen that the $N^{-1/2}$ scaling of energy gaps for $V=0$
phase boundary arises from the
same mathematical origin as in the case of the stochastic line. The scaling
amplitudes for general levels are given in terms of a scaling variable
$u\sim (H-H_c)N$ where $H_c$ is the position of the phase boundary at
$V=0$.  Their crossover behaviors enable one to identify
the class of levels which become degenerate at the phase boundary of 
$|H| < H_c$.
At $H=H_c$, the energy gap amplitudes are found to be proportional to 
a sum of square-root of integers and an anomaly term.
Furthermore, the $O(1)$ part of the partition function in the 
scaling limit, which is simply
called the partition function in this work, 
is also obtained for both the chain and the lattice model 
in terms of several infinite products involving non-integer
powers of nome.
Similar analysis is also carried out at the phase boundary of $H=0$. Here,
the energy gaps scale as $N^{-2}$ which is a characteristic of the
PT transition. Also  the corresponding 
partition functions are evaluated and 
shown to be that of the  non-relativistic free 
fermion system with dispersion relation $\varepsilon \sim k^2$.
 Using
physical requirement of invariance of the partition function of the
lattice model under exchange of rows and columns, i.e. under $\pi/2$
rotation of the lattice, we find interesting
mathematical identities between infinite products. 
We also argue that the $N^{-1/2}$ scaling at the
$V=0$  phase boundary is a feature arising from viewing the standard
PT transition of the $H=0$ phase boundary  
 after $\pi/2$ rotation.

This paper is organized as follows. In Section \ref{sec:bethe}, we set
notations and review
the Bethe ansatz method for bulk properties. Then we  discuss
classifications of low-lying excitations. In Section \ref{sec:energy}, 
the formalism for calculating finite-size corrections in energy
is developed in line with I and the leading order solution 
is expressed in terms of level-dependent scaling functions
with the scaling variable $u$. In Section
\ref{sec:crossover}, the crossover behaviors of the energy gaps for
$H > H_c$ and $H < H_c$ are discussed. From this, the class of excitations
which remain massless at the facet boundary of $|H| <H_c$  is identified.
In Section \ref{sec:spectra}, the spectra and the partition
functions at $H=H_c$ are derived. 
We discuss in Section \ref{sec:spectra2}, 
the spectra and the partition
functions  at the antiferromagnetic phase boundary of $H=0$.
 From this, we obtain several identities between infinite
products.  Finally, we summarize and discuss our results in Section
\ref{sec:discuss}. 
Some mathematical details are relegated to Appendices.
Appendix \ref{sec:recursion} proves some properties of the scaling function,
Appendix \ref{sec:partition}
 shows the details of derivations for the partition
function at the $V=0$ phase boundary,
while Appendices \ref{sec:spectra3} and  \ref{sec:partition2}
 derive the spectra and the partition
function, respectively,  at the $H=0$ phase boundary.

 \section{Bethe ansatz and energy levels}
 \label{sec:bethe}

 We consider the asymmetric XXZ chain of $N$ sites whose Hamiltonian is
given by 
\be   \label{ham}
{\cal H} =(2 \sinh \lambda)^{-1}
  \sum_{i=1}^N \{ \frac{\cosh \lambda}{2}(\sigma_i^z \sigma_{i+1}^z
  -1 ) - e^{2H}\sigma_i^+\sigma_{i+1}^- 
      - e^{-2H}\sigma_i^-\sigma_{i+1}^+ \} 
\ee
where $\sigma_i^a$ are the Pauli spin operators,
$\sigma_{N+1}^a=\sigma_1^a$, and $H$ is the horizontal field. The standard
interaction parameter $\Delta$ has been parametrized as $\Delta = -\cosh \lambda$
 with $ \lambda > 0$ since 
we are interested in the region $\Delta< -1$. The front factor is included for
later conveniences.
 Below we use the short notation 
\be  
\alpha = \exp (-2 \lambda).
\ee 
We also use the notation for the magnetization as 
\be \label{Q}
\sum_{i=1}^N \sigma_i^z =  N-2Q = 2 r
\ee
where $Q$ is the number of down spins and use 
\be \label{q}
 q = Q/N = \frac{1}{2} - \frac{r}{N}
\ee
so that $0\leq q \leq 1$ and $r$ takes the integer (half-integer) values
for $N$ even (odd). We work in sectors of general but finite $r$  as
$N \rightarrow \infty$. Since the spectra of Eq.\ (\ref{ham}) is symmetric with
respect to $H$ and $r$, we work in the region of $H \geq 0$ and $r \geq 0$. 
When a vertical field $V$ is present, it contributes an additional energy
$-2rV$ in Eq.\ (\ref{ham}) but we set $V=0$ except in Section
\ref{sec:spectra2} where the spectra of the $H=0$ phase boundary are
discussed. 

The eigenvalues $E$ of $\cal H$ in the sector $Q$ are given by 
\be \label{e}
E =  \sum_{j=1}^{Q} f_E (x_j) 
\ee
where 
\be \label{fe}
f_E (x) = -\frac{x}{x-1}- \frac{ \alpha x}{1-\alpha x},
\ee
and $\{x_j\}$ are the solution of the Bethe ansatz equation

\begin{equation} \label{bae}
\left( e^{2H-\lambda} \frac{x_i-1}{1-\alpha x_i} \right)^N =
 (-1)^{Q-1} \prod_{j=1}^Q \frac{x_i -\alpha x_j}{x_j -\alpha x_i}
  \qquad (i=1,2,\ldots,Q).  \end{equation}
(The roots $\{x_j\}$ are related to the $\{ \alpha_j \}$ 
notation of \cite{albdw,liebwu,nolden}
by  $x_j = \exp ( \lambda - i \alpha_j )$. )

 The asymmetric XXZ Hamiltonian is the logarithmic derivative of 
the asymmetric 6-vertex model row-to-row transfer matrix 
at the extreme anisotropic limit \cite{kim95}.
 The eigenvalues
$ \Lambda $
of the latter are also expressed in terms of $\{ x_j \}$. In this work,
we consider the 
isotropic 6-vertex model, or the F model, 
in external fields  whose six Boltzmann weights are given
as $\omega_{1,2}=\exp (\pm(H+V))$, $\omega_{3,4}=\exp (\pm (H-V))$ and
$\omega_{5,6} = 2 \cosh (\lambda /2)$, respectively. The transfer matrix
eigenvalues of  the F model are  given as
\be \label{loglam}
  -\ln \Lambda =- HN- 2 r V - \sum_{j=1}^Q f_{F} (x_j) 
\ee
where 
\be \label{ff}
f_{F} (x) = \ln \frac{x- \exp (-\lambda)}{1-\exp (-\lambda) x}
\ee 
for $H >0 $. Following discussions can be easily extended 
to the anisotropic 6-vertex
model but we consider only the isotropic case for simplicity.
We define the phase function or the counting function $Z_N (x)$ by
\be \label{zn}
 i Z_N (x) = 2 H-\lambda+(1-q) \ln x + \ln \frac{1-x^{-1}}{1-\alpha x}
 + \frac1N \sum_{j=1}^Q f_Z (x,x_j) 
\ee
with 
\be \label{fz}
f_Z (x,x') = \ln x' - \ln \frac{1- \alpha x'/x}{1- \alpha x/x'}
            =\ln x' + \sum_{n \neq 0} \frac{\alpha^{|n|}}{n}
         \left( \frac{x'}{x} \right)^n 
\ee 
and the root density function $R_N (x)$ by the derivative of $Z_N$ as
\bea \label{rn}
 R_N (x) &=&  i x Z_N ' (x) \nonumber \\
&=& -q +\frac{x}{x-1}+\frac{\alpha x}{1-\alpha x}
 + \frac{1}{N} \sum_{j=1}^Q f_R (x,x_j) 
\eea
where  
\be \label{fr}
 f_R (x,x') = - \sum_{n \neq 0} \alpha^{|n|} \left( \frac{x'}{x}
 \right)^n 
\ee
with $\sum_{n \neq 0}$ denoting  the sum over all integers except 0. 
In terms of $Z_N$, the Bethe ansatz equation can be rewritten as
\be \label{bae2}
  Z_N (x_j) = \frac{2 \pi}{N} I_j
\ee
where $I_j$ are half-integers (integers) for $Q$ even (odd).
The ground state for each $Q$ is obtained if one chooses $I_j$ as
\be \label{ij}
\frac{2 \pi}{N} I_j = \frac{2 \pi}{N} \left( -\frac{Q+1}{2} +j \right)
\equiv \phi_j,
\ee
for $j$=1,2,$\ldots$,$Q$.
In the thermodynamic limit, $\{ x_j\}$ for the ground state 
form a continuous curve
given by the locus $x=Z_\infty^{-1} (\phi)$ for $-\pi q \leq \phi
 \leq \pi q$ where
 $Z_N^{-1} (\phi)$ denotes the inverse function of $Z_N(x)$. 

The simplifying feature for the case of $q=1/2$ and $ |H| \leq H_c$ 
 is that the root contour becomes closed in the $x$-plane as for 
 the case of the stochastic line.
Here, $H_c$ is the antiferromagnetic transition point to be given in
Eq.\ (\ref{hc}). To find actual form of $Z_\infty(x)$ for $q=1/2$ and
$|H| < H_c$,
one may evaluate the sum in Eq.\ (\ref{zn})
by a contour integration over the circle $|x| = \exp(\lambda - b)$ with
$ -\lambda < b < \lambda $  
using the known expression $R_\infty (x)$ which is 
\be \label{rinfty}
R_\infty (x) = \sum_{n} \frac{\alpha^n}{ 1+\alpha^n} x^n 
\ee
with $\sum_n$ denoting 
 the sum  over all integers and $1 < |x| < 1/\alpha$.
The result is
\be \label{zinfty}
i Z_\infty (x) = 2H-\lambda +\frac12 \ln x +\frac12 \ln |x_0| +
 \sum_{n \neq 0} \frac{\alpha^n}{
 n (1+\alpha^n)} (x^n + x_0^n)
\ee
where  $x_0 \equiv \exp ( \lambda - b + \pi i) $ is the end point of
the contour. The conditions $Z_\infty ( \exp (\lambda -b  \pm
\pi i)) =\pm \pi /2$ are  met if $H$ is related to $x_0$ or $b$ by
\be \label{2h}
2 H = \lambda - \ln |x_0| - 2 \sum_{n \neq 0} \frac{\alpha^n}{
 n (1+\alpha^n)} x_0^n.
\ee
It is convenient to use the real function $\Xi (b)$ defined by 
 \cite{liebwu,nolden}
\bea \label{xib}
\Xi (b)&=& \frac{b}{2} + \sum_{n=1}^\infty 
\frac{(-1)^n \sinh n b }{n \cosh n \lambda} \nonumber
\\ &=& \ln \frac{\cosh(\lambda+b)/2}{\cosh (\lambda -b)/2} -
   \frac{b}{2} -\sum_{n=1}^\infty \frac{(-1)^n \alpha^n \sinh n b}{
  n \cosh n \lambda}.
\eea
Eq.\ (\ref{2h}) is then the same as
\be \label{h}
H = \Xi (b)
\ee
while $R_\infty (x_0) = \Xi ' (b)$ with ' denoting the derivative with
respect to $b$. 

As $b$ varies from 0 to $\lambda$, $x_0$ moves from $-e^\lambda$ to $-1$
and $H$ increases from 0 to 
\be \label{hc}
 H_c = \Xi (\lambda) .
\ee
It turns out that  the critical value $H_c$ 
  is indeed the value of the antiferromagnetic phase boundary
at $V=0$. Moreover the entire antiferromagnetic 
phase boundary of the lattice model is described by the $H$-dependent
critical vertical field $V_c$ given by  \cite{suthyy}
\be \label{vclattice} 
 V_c =  \Xi (\lambda -b) \makebox[3cm ]{\ } \mbox{(lattice)}
\ee 
with $-2 \lambda<b<2 \lambda$. 
That of the asymmetric XXZ chain with normalization as given in 
Eq.\ (\ref{ham}) is given by (See Appendix C)
\be \label{vcchain}
 V_c = \Xi' (b)  \makebox[3cm ]{\ } \mbox{(chain)} .
\ee
Typical phase diagrams in the $H$-$V$ plane are shown in Fig. 1.

The bulk ground state energy per site for $|H| \leq H_c$ is given 
as $e_\infty \equiv \lim_{N\rightarrow\infty} E/N = 
-  e_0 $ where 
\be \label{e0}
 e_0 = \frac12 + 2 \sum_{n=1}^\infty \frac{\alpha^n}{1+\alpha^n}
\ee
and the bulk free energy of the F model is
$ f_\infty \equiv   \lim_{N\rightarrow \infty} (-\ln \Lambda)/N =
 - \lambda/2 - \sum_{n=1}^\infty n^{-1}\alpha^{n/2}\tanh n\lambda 
$.  

In the antiferromagnetic phase, finite mass gaps appear in
the energy spectra. But these mass gaps are offset by fields and 
at the antiferromagnetic phase boundary, infinite number of other
levels become degenerate with the ground state in the thermodynamic
limit. These degeneracies are lifted for finite but large $N$.
Our aim here is to find these finite-size corrections 
for arbitrary energy levels. We denote these energy gaps
 as $\Delta E$ and $\Delta F$,
respectively:
 $ \Delta E = E - e_\infty N $ and $
  \Delta F = -\ln \Lambda - f_\infty N$ .

The ground state characterized by Eq. (\ref{ij}) corresponds to a Fermi
sea as depicted in Fig.\ 2(a) for $r$=0.
 Other energy levels are obtained if other sets of $\{ I_j \}$ are
chosen. An important class of levels is the $m$-shifted levels where
$\{ I_j\}$ are shifted by an integer $m$ from that of the ground state, 
i.e. $2\pi I_j /N = \phi_{j+m}$. These states are denoted as ($r$,$m$,0).
Fig.\ 2(b) shows the 1-shifted level (0,1,0).
Further excitations are obtained by creating from the $m$-shifted states
an equal numbers of particles and holes near the two ends of the Fermi 
sea. Excited levels are then  characterized by
their particle and  hole positions. 
 The particle (hole) positions near the right-hand side of the Fermi sea
can be specified by a set of  integers $\{p_k\}$, ($\{h_k\}$) which are 
related to the index  
$j$ in $\phi_j$ as $j=Q+m+p_k$ ($Q+m+1-h_k$), for 
$k$=1,2,$\ldots$,$n_+$, $n_+$ being the number of particle-hole pairs, and
 $1\leq p_1 < p_2 < \cdots < p_{n_+}$
($1\leq h_1 < h_2 < \cdots < h_{n_+}$).
Particles (holes)  near the left-hand side may be labeled similarly by
$\{ \bar{p}_k\}$ ( $\{ \bar{h}_k\}$) with 
$j=1+m-\bar{p}_k$ ($m+\bar{h}_k$),
 for $k$=1,2,$\ldots$,$n_-$ and $1\leq \bar{p}_1 < \bar{p}_2 < \cdots < 
\bar{p}_{n_-}$ ($1\leq \bar{h}_1 < \bar{h}_2 < \cdots <
\bar{h}_{n_-}$)
. Such excitation configurations are denoted
collectively by $\cal P$. General levels are then labeled as ($r$,$m$,
$\cal P$). Fig.\ 2(c) shows a level obtained by 
creating $n_+$=2 particle-hole pairs at
the right-end and $n_-$=1 pair
 at the left-end from the 1-shifted level of 
Fig.\ 2(b). 
 In \cite{nk3}, it was shown for the critical region
that levels obtained in this way
account for all the low-lying excitations 
 which are expected to appear from the central charge 
$c=1$ conformal field theory. Here, we also assume that all the low-lying
excitations are generated in this manner. 

When the particle-hole configurations differ from the ground state
only at finite distances away from the two ends of the Fermi sea, 
finite-size corrections in energies are determined by the analytic
property of $Z_\infty (x)$ near $x=x_0$. However, the root density
vanishes at $x=x_c^0 \equiv  e^{\pi i}$ 
which is the end point of the root contour
for $H= H_c$. Therefore $Z_\infty^{-1} (\phi)$ exhibits a square root
singularity at $\phi = \pm \pi/2$. The same phenomenon was the 
origin of the unusual $N^{-3/2}$ scaling of the energy gaps along
the stochastic line and 
one can expect similar mechanism produces $N^{-1/2}$ scaling in the
present case. On the other hand, for $|H| < H_c$ , $O(N^{-1})$ variations
in $\phi$ near $\phi=\pm \pi/2$ cause, via $Z_\infty^{-1} (\phi)$,
$O(N^{-1})$ variations in $x$-plane and hence
the finite-size corrections take the form of power series in $N^{-1}$.
What is special at the antiferromagnetic phase boundary is that the real parts
of the energy gaps 
scale as $N^{-2}$ which is the characteristic of the PT 
transition. These features are born out explicitly below. 

\section{Energy gaps for $H$ near $H_c$}
\label{sec:energy}
   
In this section, we derive the leading order finite-size correction
of the energy gaps $\Delta E$ and $\Delta F$ for general levels
with $(H-H_c)N$ as a scaling variable.
Readers not interested in details may skip to Eqs.\ (\ref{dele}) and 
(\ref{delf}) which are the main results of this section. 

To find the energy gaps for $H$ near $H_c$, we employ the method of I.
Thus, for finite but large $N$ and for levels whose energies are close to that
of the ground state, we assume, following I, that there exists in the
complex $x$-plane a point $x_c$ near $x_c^0=e^{\pi i}$ 
such that $Z_N'(x_c)=0$, $Z_N(x)$
itself having a branch cut passing through $x_c$,  and that near
$x_c$, $Z_N (x)$ has the expansion of the form
\be \label{zn2}
i Z_N (x) = \pm i \pi q  + y_{\pm} \frac{\pi}{N} + \frac{1}{a_1^2}
 (x-x_c)^2 + O((x-x_c)^3) 
\ee
where the upper (lower) sign refers to the assumed
value of $i Z_N(x)$ at $x=x_c$ ($x=x_c e^{-2\pi i}$). 
This amounts to assuming 
\be \label{zninv}
Z_N^{-1} (\pm \pi q + \frac{\pi}{N} \xi) =
x_c^\pm \pm i a_1 \sqrt{ y_\pm -i \xi}  \sqrt{\frac{\pi}{N}} + O(N^{-1})
\ee 
for $\xi \sim O(1)$ with $x_c^+ =x_c$ and $x_c^-=x_c e^{-2\pi i}$. 

The constants $x_c$, $a_1$, $y_\pm$, etc. are
all level dependent and should be determined self-consistently.
But, first we assume they are known and evaluate the finite sums  
of the form
\be \label{sf}
S[ f ] = \sum_{j=1}^Q f(x_j) = \sum_{j=1}^Q f( Z_N^{-1} (\frac{2\pi}{N}
I_j)) 
\ee
to $O(1/\sqrt{N})$ for $f(x')$ given in Eqs. (\ref{fe}), (\ref{ff}),
(\ref{fz}) and (\ref{fr}).
Following the same steps as in I, but generalizing to levels
with $m \neq 0$, we find 
\bea \label{sf2}
S[ f] &=& \frac{N}{2\pi i} \oint f(x) R_N(x)  \frac{dx}{x} +
 (m+\frac{i}{2} y_+) f(x_c) - (m+\frac{i}{2} y_-) f(x_c e^{-2\pi i})
\nonumber \\
& & +a_1 f'(x_c) Y_1(y_+,y_-)
 \sqrt{\frac{\pi}{N}}  + O( N^{-1} )
\eea
where the integral is over a closed contour in the annulus $1<|x|<1/\alpha$
and $Y_1 (y_+,y_-)$ is defined below.  
To define $Y_{1}$
, we first introduce two functions  $J_+ (y)$ and $J_- (y)$ of 
complex variable defined for 
$\Im y > 0$ and $\Im y <0$, respectively, as
\be \label{jy}
J_\pm (y) = \frac{1}{2} \int_0^\infty \frac{ \sqrt{y+t}-\sqrt{y-t}}{
e^{\pi t}-1} dt\pm \frac{i}{2} y^{1/2} - \frac13 y^{3/2} 
\ee
with the branch cut of the square roots at the negative real axis.
As shown in Appendix \ref{sec:recursion},
 $J_\pm(y)$ satisfy the recursion relations
\be \label{jy2}
 J_\pm (y\pm 2i) = J_\pm (y) \mp i  \sqrt{y} .
\ee
Thus, we may extend the definition of $J_+(y)$ ($J_-(y)$)  into the half plane
$\Im y \leq 0$ ($\Im y \geq 0$) using the relation Eq. (\ref{jy2}) recursively.
Having defined $J_\pm (y)$, we then give the expression of $Y_1$
for general level $(r,m,{\cal P})$: 
\bea \label{y1}
Y_1 (y_+,y_- ) &=&  J_+ ( y_+ + i -2im )   
                +  J_- (  y_- - i -2im ) \nonumber \\
 &\ & + i \sum_{k=1}^{n_+} \left( \sqrt{y_+ +i(1-2m-2p_k)}
-\sqrt{y_+ -i(1+2m-2 h_k)} \right) \nonumber \\
 &\ & - i \sum_{k=1}^{n_-} \left( \sqrt{y_- -
 i(1+2m-2\bar{p}_k)} -\sqrt{y_- +i(1-2m-2 \bar{h}_k)} \right).
\eea
The first two terms in Eq.\ (\ref{y1})
are contributions for the $m$-shifted
levels and the sums in the second (third) line accounts for 
particle-hole pairs created at the right- (left-) end of the Fermi sea.
 When $m=0$ and
$y_+=y_-$, $Y_1 (y,y) $ reduces to minus of $Y_1 (y)$ defined in I.

Applying Eq.\ (\ref{sf2}) to Eq.\ (\ref{rn}), we then obtain the
solution for $R_N$ as
\be \label{rn2}
R_N (x) = R_\infty (x)+ \frac{r}{N}+ \frac{1}{N} \sum_{n\neq0} 
\frac{\alpha^{|n|}}{1+\alpha^{|n|}} \left( \frac{x}{x_c}\right)^n
 \left( \frac{y_+-y_-}{2i} + \frac{n }{x_c} a_1 Y_1(y_+,y_-) \sqrt{
  \frac{\pi}{N}} \right) + O(N^{-2}).
\ee
Having obtained $R_N (x)$ to $O(N^{-3/2})$, we can evaluate $S[f]$ to
$O(N^{-1/2})$ for any $f(x)$. In particular, 
applying the general sum formula 
to $f(x')=f_Z(x,x')$, $Z_N (x)$ can be evaluated to $O(N^{-3/2})$.
 Evaluating the latter
at $x= x_c$ and $x=x_c \exp( -2\pi i) $, we obtain two self-consistency
equations (See Eq.\ (\ref{zn2}).)
\bea \label{yeq}
\pm i \pi q  + \frac{\pi}{N} y_\pm &=& \pm i(1-q) \pi  +
   2H-\lambda+ 2 \sum_{n\neq 0}
 \frac{\alpha^n}{1+\alpha^n} \frac{x_c^n}{n} \nonumber \\
&\ & +(1+ \frac{2r}{N}-\frac{y_+-y_-}{2Ni}) (\ln x_c - \pi i)
  + (2im-\frac{y_++y_-}{2})\frac{\pi}{N}  \nonumber \\
&\ & + 2 \pi^{1/2} e_0 a_1 x_c^{-1}
   Y_1(y_+ , y_-) N^{-3/2} + O(N^{-2}) 
\eea
where $e_0$ is given in Eq.\ (\ref{e0}).
 From Eq.\ (\ref{yeq}), one first obtains $y_+-y_- = 4ir$.
Next, evaluating $R_N(x)$ and $R_N'(x)$ at $x_c$, we obtain two more
self-consistency equations as
\bea \label{xc}
0 &=& R_\infty (x_c) + \frac{2 e_0 r}{N} + O(N^{-2}), 
\\ \label{a1}
\frac{2 x_c}{a_1^2} &=& R_\infty ' (x_c) + O(N^{-3/2})
\eea
where $y_+-y_- = 4ir$ has been used in Eq.\ (\ref{xc}).
Solving Eqs. (\ref{xc}) and (\ref{a1}) perturbatively, we have
\bea \label{xc2}
x_c &=& e^{\pi i} +  \frac{2r e_0}{e_1^2} \frac{1}{N} +O(N^{-2}),
\\ \label{a12}
 a_1 &=& \frac{\sqrt{2}}{ e_1} +O(N^{-1})
\eea
where $e_1$ is a positive real constant defined as
\be \label{e1}
e_1 \equiv (-R_\infty '(-1))^{1/2} =(-\Xi''(\lambda))^{1/2}=
         \left( 
\frac{1}{4}+ 2 \sum_{n=1}^\infty 
   n (-1)^n \frac{\alpha^n}{1+\alpha^n}\right)^{1/2} .
\ee
Using Eqs.\ (\ref{xc2}) and (\ref{a12})  in Eq. (\ref{yeq}), we find the
leading order solution of $y_\pm$ as 
\be \label{ysol}
y_\pm = (\pm 2r +m)i + u + O(N^{-1/2}) 
\ee
where $u$ is the scaling variable 
\be \label{u}
 u = (H-H_c)N /\pi 
\ee
which is assumed fixed as $N\rightarrow \infty$. 
Having determined
$y_\pm$, $x_c$ and $a_1$ to necessary orders, we are now in a position to 
 evaluate $\Delta E$ and $\Delta F$. Applying Eq.\ (\ref{sf2}) together
with Eq.\ (\ref{rn2}) to Eq.\ (\ref{fe}), we find
\bea \label{e2}
E &=&- e_0 N -r + \frac{y_+-y_-}{2i} \nonumber \\
  &\ & + \sum_{n\neq0} \frac{\min (1,
\alpha^{-n})}{1+\alpha^{|n|}}x_c^{-n} \left( \frac{y_+-y_-}{2i} 
+ n x_c^{-1} a_1 Y_1 (y_+,y_-) \sqrt{ \frac{
 \pi}{N}} \right)  +O(N^{-1}). 
\eea
Inserting the solutions Eqs.\ (\ref{xc2}), (\ref{a12}) and  (\ref{ysol}), to
Eq.\ (\ref{e2}), the $O(1)$ terms in Eq.\ (\ref{e2}) cancel out and
we finally have, for $V=0$,
\be \label{dele}
\Delta E \equiv E-e_\infty N = e_1 {\cal Y}_1^{(r,m,{\cal P})}  
\sqrt{ \frac{2\pi}{N}} +O(N^{-1})
\ee
where  $e_\infty=-e_0$, $e_1$ is given in Eq.\ (\ref{e1})
 and ${\cal Y}_1^{(r,m,{\cal P})}$ is the value
of $Y_1 (y_+,y_-)$ at $y_\pm =u+ (\pm 2r+m)i$:
\bea \label{caly}
 {\cal Y}_1^{(r,m,{\cal P})}  &=&  J_+ (u+i (2r-m+1) )
                +  J_- (u-i(2r+m+1)   ) \nonumber \\
 &\ & + i \sum_{k=1}^{n_+} \left( \sqrt{u +i(2r-m+1-2p_k)}
-\sqrt{u +i(2r-m-1+ 2 h_k)} \right) \nonumber \\
 &\ & - i \sum_{k=1}^{n_-} \left( \sqrt{u -
 i(2r+m+1-2\bar{p}_k)} -\sqrt{u -i(2r+m-1+2 \bar{h}_k)} \right).
\eea
Here, $J_\pm(y)$ and $u$ are given in Eq.\ (\ref{jy}) and (\ref{u}),
respectively.  ${\cal Y}_1^{(r,m,{\cal P})}$ regarded as a function of
the scaling variable $u$ is then the scaling function.
Similar calculation for the F model leads, for $V=0$, 
\be \label{delf}
\Delta F \equiv  -\ln \Lambda -f_\infty N = -i\pi  m +\frac{e_2}{e_1}
    {\cal Y}_1^{(r,m,{\cal P})}
\sqrt{ \frac{2\pi}{N}} +O(N^{-1}) 
\ee
where the constant $e_2$ is given as 
\be \label{gamma}
 e_2 \equiv \Xi '(0)=
   \frac12 + \sum_{n=1}^\infty \frac{(-1)^n}{ \cosh n\lambda  }.
\ee
Note that  $\Delta F$ 
 has an imaginary term due to $\Lambda$ being negative for $m$ odd.
 Apart from this, all energy gaps scale as $N^{-1/2}$ and all the
level dependences are encoded in the scaling function
 $ {\cal Y}_1^{(r,m,{\cal P})}$. 
 Higher order corrections which come as a power series in $N^{-1/2}$
 can be calculated 
perturbatively using the method similar to that of I.

\section{Crossover behaviors for $H \neq H_c$}
\label{sec:crossover}

Eqs.\ (\ref{dele}) and (\ref{delf})
 are obtained with the scaling variable $u$
fixed. By considering the limits $u \rightarrow \pm \infty$, we can derive how
the spectra crossover to those of the critical and massive phases.
First consider the case 
$H>H_c$ where  $u\rightarrow \infty$ as $N\rightarrow \infty$. From Eq.\
(\ref{jy}), one can show that, as $u\rightarrow \infty$, 
\be \label{ju1}
J_\pm (u\pm i a) = -\frac{1}{3} u^{3/2} \mp \frac{i}{2} (a-1) u^{1/2}
 + (\frac{a^2}{8}-\frac{a}{4}+ \frac{1}{12}) u^{-1/2} + O(u^{-3/2})
. \ee 
Using this in Eq.\ (\ref{caly}),  ${\cal Y}_1^{(r,m,{\cal P})}$ becomes, 
\be \label{caly2}
{\cal Y}_1^{(r,m,{\cal P})} = -\frac{2}{3} u^{3/2} + imu^{1/2}+
(\frac{m^2}{4} +r^2 - \frac{1}{12} + {\cal N} + \overline{\cal N} )
u^{-1/2} +O(u^{-3/2})
\ee 
with ${\cal N}= \sum_{k=1}^{n_+} ( {p}_k+ {h}_k-1)$ and 
$\overline{\cal N}= \sum_{k=1}^{n_-} ( \bar{p}_k+ \bar{h}_k-1)$.
The first term in Eq.\ (\ref{caly2}) contributes a bulk energy term  
proportional to $(H-H_c)^{3/2}N$ to $\Delta E$ and $\Delta F$, the second an
$O(1)$ imaginary term and the
 third a real one proportional to $(H-H_c)^{-1/2}N^{-1}$.
These are exactly those which appear in the critical phase as shown in 
\cite{nk3} and show the complete 
operator content of the Gaussian model. In the Coulomb gas
picture, $m$ and $r$ are the spin wave and vortex quantum numbers,
respectively, 
characterizing primary operators of the $c=1$ conformal field theory whose
Gaussian coupling constant $g$ is 2,
and ${\cal N}$ and $\overline{\cal N}$ account for the conformal
towers generated from the primary operators. 
Higher order corrections
in Eqs. (\ref{dele}) and (\ref{delf}) not discussed here would give terms
which are  higher orders in $(H-H_c)$. 

Next consider the case $H<H_c$. Now $u\rightarrow -\infty$ as $N \rightarrow 
\infty$ and one has to be careful about the branch cut of the square root. 
We find
\be \label{ju2}
J_\pm (u \pm i a)= \pm \frac{i}{3} |u|^{3/2} + G_\pm |u|^{1/2} + O(|u|^{-1/2})
\ee
with $G_\pm = |a-1|/2$ for $a>0$ or for $a$ negative odd integers 
but, for $a \leq 0$ and even integers, $G_+ = (|a|-1)/2$ while $G_- = 
 (|a|+3)/2$. Using this,
 we find the leading order of  $ {\cal Y}_1^{(r,m,{\cal P})}$
for levels ($r$,$m$,0) as 
 $ {\cal Y}_1^{(r,m,0)} = \max (2r, |m|) |u|^{1/2}$ if $(2r+m)$ is even and
  $ {\cal Y}_1^{(r,m,0)} = \max (2r, |m-1|) |u|^{1/2}$ if $(2r+m)$ is odd. 
Thus,
 when $m$ is in the range $-(2r+1)< m \leq (2r+1)$, $ {\cal Y}_1^{(r,m,0)}
=2r |u|^{1/2} +O(|u|^{-1/2}) $. As to the contributions from particle-hole
excitations in Eq.\ (\ref{caly}), we note that, for $-(2r+1)< m \leq
(2r+1)$,  
particles  at positions satisfying the conditions $2r-m+1-2p_k \geq 0$ or 
$2r+m+1-2 \bar{p}_k > 0$ contribute $-|u|^{1/2}$
but those violating the conditions
and all holes contribute $|u|^{1/2}$. Thus when all particle positions 
satisfy the conditions, or equivalently, when the last positions satisfy
the condition
\be \label{cond}
-(2r+1)<m\leq (2r+1),  \makebox[1cm]{\ } 2p_{n_+}
 \leq 2r-m+1, \makebox[1cm]{\ } 2\bar{p}_{n_-} < 2r+m+1 ,
\ee 
then  all particle contributions are 
cancelled by 
holes.  Thus, for such levels, $ {\cal Y}_1^{(r,m,{\cal P})}
=2r |u|^{1/2} +O(|u|^{-1/2}) $ and $\Delta E$ becomes
\be \label{egap}
   \Delta E = 2 \sqrt{2} e_1 (H_c-H)^{1/2} r
\ee 
as $H\rightarrow H_c$  in the thermodynamic limit. This is the mass gap of
the antiferromagnetic phase of the asymmetric XXZ chain.
The mass gap for general $H$ is derived in
\cite{albdw} for levels $(r,0,0)$  and is in agreement 
with Eq.\ (\ref{egap}). 
It is also derived for general $H$ implicitly in Eq.\ (\ref{dele4}) and
is in accord with Eq.\ (\ref{vcchain}).
Corresponding mass gap for the lattice model is
\be \label{fgap}
 \Delta F = -i\pi m + 2 \sqrt{2}e_2 e_1^{-1} (H_c-H)^{1/2} r.
\ee 
For levels which do not satisfy Eq.\ (\ref{cond}),
the mass gap becomes higher by $O(1)$. 
Eqs.\ (\ref{egap}) and (\ref{fgap}) are valid for $r\geq 0$; they are
valid for general $r$ if  $r$ is replaced by $|r|$.
When the vertical field $V$ is applied, both the chain and lattice model
energy acquire a trivial term $-2rV$. When $2rV$ is equal to the mass gap
given in Eq.\ (\ref{egap}) and (\ref{fgap}) for the chain and lattice model,
respectively,
all levels $(r,m,{\cal P})$  satisfying Eq.\ (\ref{cond})
and $r\geq0$ become degenerate with the ground state energy in the
thermodynamic limit.  Finite-size corrections for these states will be 
discussed in Section \ref{sec:spectra2}.

A special mention is needed for $r=0$ which is the half-filled sector
for $N$ even. Here, only two levels $(0,0,0)$ and $(0,1,0)$ satisfy Eq.\
(\ref{cond}). The latter which has momemtum $\pi$ is the exponentially
degenerate second largest eigenvalue of the transfer matrix as discussed
in \cite{baxter73}. Such exponential degeneracy is reflected in the
large $u$ behavour of   $ {\cal Y}_1^{(r,m,{\cal P})}$. When expanded as a
power series in $|u|^{-1}$, both ${\cal Y}_1^{(0,0,0)}$ and ${\cal
Y}_1^{(0,1,0)}$ have vanishing coefficients to all orders in the series.

\section{Spectra  and partition functions at $H_c$}
\label{sec:spectra}

Now we consider the case $H=H_c$ (The points A in Fig. 1).
Since $u=0$ in this case, one needs to evaluate
$J_\pm(ia)$ for integer values of $a$. Successive application of
the recursion relation Eq.\ (\ref{jy2}) yields, 
\be \label{jp}
J_+(ia) = \left\{ 
  \begin{array}{ll}
    \sqrt{-i} \sum_{j=0}^{(a-3)/2} \sqrt{1+2j}-\sqrt{-i}c_1 & \mbox{if
 $a > 0$ and odd} \\
  \sqrt{i} \sum_{j=0}^{(|a|-1)/2} \sqrt{1+2j}-\sqrt{-i}c_1 & \mbox{if
 $a <0$ and odd} \\
\sqrt{-i} \sum_{j=0}^{(a-2)/2} \sqrt{2j}+\sqrt{-i}c_2 & \mbox{if
 $a > 0$ and even} \\
\sqrt{i} \sum_{j=0}^{|a|/2} \sqrt{2j}+\sqrt{-i}c_2 & \mbox{if
 $a \leq 0$ and even} 
   \end{array}  \right. 
\ee
and $J_-(-ia)=J_+(ia)^*$ where the positive numbers $c_1$ and $c_2$,
called anomaly, are given as
\bea \label{c1}
c_1 &\equiv& -J_+(i)/\sqrt{-i} = \frac{1}{6}-\frac{1}{2i} \int_0^\infty
\frac{\sqrt{1+it}-\sqrt{1-it}}{e^{\pi t}-1} dt 
 \\ \label{c2}
c_2 &\equiv& J_+(0)/\sqrt{-i} = \zeta(3/2)/(2\sqrt{2}\pi)=0.293\,995\,52\ldots
\eea
with  $\zeta(z)$  the Riemann zeta function.
We will see later in an indirect way that $c_1=(\sqrt{2}-1)\zeta(3/2)/
(4\pi)=0.086\,109\,29\ldots$ (Eq.\ (\ref{c12})).

 Using Eq.\ (\ref{jp}) in Eq.\ (\ref{caly}) with $u=0$,
 one can read off
 $ {\cal Y}_1^{(r,m,{\cal P})}$ for arbitrary levels. It is a sum of 
square-root of integers and the anomaly term.  We give a few
examples:

(i) Ground state in each sector $r$ for $N$ even: 
\be \label{r00}
 {\cal Y}_1^{(r,0,0)} = 
  \sqrt{2} \left( \sum_{j=0}^{r-1} \sqrt{1+2j}
   -c_1 \right) .
\ee

(ii) $m$-shifted levels in sector $r=0$:
\be  \label{mshift}
 {\cal Y}_1^{(0,m,0)} = \left\{
 \begin{array}{ll}
  2\sqrt{i}(1+\sqrt{3}+\cdots+\sqrt{m-1})-\sqrt{2}c_1 & \mbox{ for $m>0$
even} \\
 2\sqrt{i}(\sqrt{2}+\sqrt{4}+\cdots+\sqrt{m-1})+\sqrt{2} c_2 &
  \mbox{ for $m>0$ odd.} \end{array} \right. 
\ee
For $m<0$, ${\cal Y}_1^{(0,m,0)}$ is the complex conjugate of ${\cal
Y}_1^{(0,|m|,0)}$.

(iii) $m$-shifted levels with $-(2r+1)<m\leq (2r+1)$:
\be \label{rm0e}
{\cal Y}_1^{(r,m,0)}  = -\sqrt{2}c_1+\sqrt{-i} \sum_{j=0}^{r-m/2-1}
 \sqrt{1+2j}+ \sqrt{i} \sum_{j=0}^{r+m/2-1}\sqrt{1+2j}
\ee 
for $(2r+m)$ even and
\be \label{rm0o}
{\cal Y}_1^{(r,m,0)}  = \sqrt{2}c_2+\sqrt{-i} \sum_{j=0}^{r-(m+1)/2}
 \sqrt{2j}+ \sqrt{i} \sum_{j=0}^{r+(m-1)/2}\sqrt{2j}
\ee
for $(2r+m)$ odd.

(iv) One particle-hole pair excitation from (0,0,0): 
\be \label{ph}
   {\cal Y}_1^{(0,0,{\cal P})} = -\sqrt{2}c_1+ \sqrt{i} \sqrt{2p_1-1}
           +\sqrt{-i}  \sqrt{2h_1 -1}
\ee
with the positive integers $p_1$ ($h_1$) being the position of the particle
(hole).

We have compared above predictions with existing numerical data of
  $\Delta E$ obtained by solving the Bethe ansatz equation 
for finite $N$ 
\cite{wehd,albdw}. Extrapolations of available data 
($N \leq 80$ for levels (0,0,0) and (1,0,0), $N \leq 68$ for
(0,2,0), $N\leq 56$ for (2,0,0)) all show  excellent convergence to
the expected exact values.

Having obtained the leading finite-size corrections for arbitrary levels,
one can go further to evaluate the partition functions in the scaling
limit. We consider the
partition function of the asymmetric XXZ chain at temperature $T$
(in units of the Boltzmann constant)  and the lattice model on
$N \times M$ lattice with periodic boundary conditions 
 defined by
${\cal Z}_{E} = \sum \exp ( - \Delta E/T)$
and ${\cal Z}_{F} = \sum \exp ( -M \Delta F)$, respectively,
 where the sums are over all states and
the bulk part $\exp (-Ne_\infty /T)$ for the chain and $\exp (-MNf_\infty)$
for the lattice model  has been taken out. To obtain a
non-trivial limit, we take the thermodynamic limit with
$\sqrt{N}T$ or $MN^{-1/2}$ fixed. 
In this scaling limit, the $O(N^{-1/2})$ part of the energy gaps 
produces finite contributions while levels which are not 
degenerate  with the ground state in the thermodynamic limit need not
be considered.

Considering  ${\cal Z}_F$ first,  and we can write from Eq.\
(\ref{delf}), 
\bea \label{zf}
{\cal Z}_F  &=& \sum \exp ( - M \Delta F) \nonumber \\
 &=& \sum_{(r,m,{\cal P})} (-1)^{mM} \exp ( - \frac{e_2 M }{e_1}
\sqrt{\frac{2\pi}{N}} {\cal Y}_1^{(r,m,{\cal P})} ).
\eea
The sum is conveniently expressed in terms of a complex parameter
${\sf q}$ which
we call nome:  
\be \label{sfq}
{\sf q}  = \exp ( - \pi \sqrt{i \tau})
\ee
where
\be \label{tau}
\tau = 2e_2^2 M^2 /(\pi e_1^2 N).
\ee
Derivation of the partition function is relegated to Appendix
\ref{sec:partition}. Our final result for ${\cal Z}_F$ is
\be \label{zf2}
{\cal Z}_F ( {\sf q}) = \left\{ \begin{array}{ll}
 {\cal Z}_1 ({\sf q}) \pm {\cal Z}_2 ({\sf q})+ {\cal Z}_3
({\sf q}) & \mbox{ for $M$ even} \\
 \pm {\cal Z}_1 ({\sf q}) + {\cal Z}_2 ({\sf q}) \mp {\cal Z}_3
({\sf q}) & \mbox{ for $M$ odd} \end{array} \right. 
\ee
where the upper (lower) sign is for $N$ even (odd) and ${\cal Z}_i({\sf
q})$ are defined as
\bea \label{z1}
 {\cal Z}_1 ({\sf q})&=&\frac{1}{2} ({\sf q}{\bar{\sf q}})^{-c_1}
   \prod_{j=-\infty}^\infty \left( 1+ {\sf q}^{\sqrt{1+2j}} \right)
                    \left( 1+ \bar{\sf q}^{\sqrt{1+2j}} \right) \\
\label{z2}
 {\cal Z}_2 ({\sf q})&=&\frac{1}{2} ({\sf q}{\bar{\sf q}})^{-c_1}
   \prod_{j=-\infty}^\infty \left( 1- {\sf q}^{\sqrt{1+2j}} \right)
                    \left( 1- \bar{\sf q}^{\sqrt{1+2j}} \right)
\\  \label{z3}
 {\cal Z}_3 ({\sf q})&=&\frac12 ({\sf q}{\bar{\sf q}})^{c_2}
   \prod_{j=-\infty}^\infty \left( 1+ {\sf q}^{\sqrt{2j}} \right)
                    \left( 1+ \bar{\sf q}^{\sqrt{2j}} \right)
\eea 
with $c_1$ and $c_2$ given in Eqs.\ (\ref{c1}) and (\ref{c2}),
respectively, and $\bar{\sf q}$ denoting the complex conjugate of 
${\sf q}$.

For the asymmetric XXZ chain, we have
\be \label{ze}
{\cal Z}_{E} = \sum_{(r,m,{\cal P})} \exp ( - \frac{e_1}{T}
\sqrt{\frac{2\pi}{N}} {\cal Y}_1^{(r,m,{\cal P})} ) .
\ee 
In this case, we can simply use the result of ${\cal Z}_F$ for $M$ even
by redefining the nome  as
\be \label{qsf1}
{\sf q}' = \exp ( - \frac{e_1}{T}
\sqrt{\frac{2\pi i}{N}}  ).
\ee
 Using this, we then have 
\be \label{ze2}
{\cal Z}_E = {\cal Z}_1 ({\sf q}') \pm {\cal Z}_2 ({\sf q}')+ {\cal Z}_3
({\sf q}')
\ee
where the $+$ ($-$) sign is for $N$ even (odd).

\section{Spectra and partition functions at $H=0$ and $V=V_c$}
\label{sec:spectra2}

The partition function of the isotropic 
 lattice model on $N \times M$ lattice
should be invariant  under rotation of the lattice by 90$^0$.
After the rotation, roles of $H$ and $V$ are also interchanged. Thus we
expect an invariance of the partition functions under $N \leftrightarrow M$
and $H \leftrightarrow V$. When the system is in  the critical  phase, this
symmetry is manifest by the modular covariance of the partition function.
Modular invariance plays an important role in classifying the conformal
invariant theories and in revealing their mathematical structures. Present
system is in a way a non-relativistic limit of the CFT and 
it is of interest to find the analog of the modular transformation.
For this purpose, we have also calculated the energy spectra of the two
models at $ |H| \leq H_c$ for those which are low-lying near 
$ V=V_c$ where $V_c$ is the $H$-dependent critical vertical
field. The method employed is similar to the case of $H= H_c$ and $V=0$
except that the ansatz for $Z_N^{-1}(\phi)$ takes a different form and that
only those levels satisfying the condition Eq.\ (\ref{cond}) and $r\geq0$ come
into the working.
Details of the derivations are given in Appendix 
\ref{sec:spectra3}.  To compare the result with that of the $V=0$ phase
boundary (the points A in Fig.\ 1), we consider only the $H=0$ phase
boundary (the points B in Fig.\ 1). Our
result  for the F model at $H=0$ is
\be \label{delf2}
\Delta F = -\pi i m -2r (V-H_c) + \frac{e_1^2}{2 e_2^2} 
{\cal Y}_4^{(r,m,{\cal P})} (\pi/N)^2 +O(N^{-3})
\ee
where $H_c$, $e_1$ and $e_2$ are given in Eqs.\ (\ref{hc}),
(\ref{e1}) and (\ref{gamma}), respectively, and 
${\cal Y}_4^{(r,m,{\cal P})}$ is given by 
\bea \label{caly4}
{\cal Y}_4^{(r,m,{\cal P})}&=&\frac{8}{3} r^3 -\frac{2}{3} r + 2r m^2 
\nonumber \\ & &
-\sum_{k=1}^{n_+} \left( (2r-m+1-2p_k)^2 - (2r-m-1+2h_k)^2 \right)
\nonumber \\ & &
-\sum_{k=1}^{n_-} \left( (2r+m+1-2\bar{p}_k)^2-(2r+m-1+2\bar{h}_k)^2
\right), 
\eea 
while that for the chain at $H=0$  is
\be \label{dele2}
\Delta E = 2re_2 + \frac{e_4}{2 e_2^2}
{\cal Y}_4^{(r,m,{\cal P})} (\pi/N)^2 +O(N^{-3})
\ee
where  $e_4 \equiv - \Xi '''(0)=
  \sum_{n=1}^\infty (-1)^{n+1}n^2/\cosh n \lambda $.
Note that $V_c$ at $H=0$
 is the same as $H_c$ for the lattice due to isotropy,
but  it is $e_2=\Xi '(0)$ for the chain.  For $H$ not equal to zero,
 imaginary $O(N^{-1})$
terms appear in $\Delta E$ and $\Delta F$. However, the real parts are
$O(N^{-2})$ for all $|H|<H_c$.

 From the result of the energy gaps,
 we are able to calculate the partition functions as
in the previous section. We consider the lattice case first since it
is more general. To compare with Eq.\ (\ref{zf2}), we interchange $M$ and
$N$ in Eq.\ (\ref{delf2}) and offset the mass gap by setting $V=H_c$.
Then the partition function of a $M\times N$ lattice at the phase boundary
 is 
\be \label{zf3}
{\cal Z}_F = \sum_{(r,m,{\cal P})} (-1)^{mN} \exp ( - \frac{\pi}{\tau}
   {\cal Y}_4^{(r,m,{\cal P})} )
\ee
with $\tau = 2e_2^2 M^2/(\pi e_1^2 N)$ as was given previously in Eq.\
(\ref{tau}). This quantity should be the same as that defined in Eq.\
(\ref{zf}) by symmetry. To express the partition function in a compact
way, we define the new nome as 
\be \label{sfp}
{\sf p} = \exp ( -\pi/\tau).
\ee
This is analogous to the nome corresponding to the conjugate modulus
in elliptic functions \cite{baxter}. 
After performing the sums over the levels, we find ${\cal Z}_F$ as
\be \label{zf4}
{\cal Z}_F = \left\{ 
\begin{array}{ll}
 \widetilde{\cal Z}_1 ({\sf p}) \pm \widetilde{\cal Z}_2 ({\sf p})
  + \widetilde{\cal Z}_3 ({\sf p}) & \mbox {for $M$ even} \\
\pm \widetilde{\cal Z}_1 ({\sf p}) + \widetilde{\cal Z}_2 ({\sf p})
  \mp \widetilde{\cal Z}_3 ({\sf p}) & \mbox {for $M$ odd} 
 \end{array} \right. 
\ee 
where the upper (lower) sign is for $N$ even (odd) and
$\widetilde{\cal Z}_i$ are defined as
\bea  \label{tilz1}
\widetilde{\cal Z}_1 ({\sf p}) &=& \frac{1}{2}\prod_{j=-\infty}^\infty
        (1+{\sf p}^{(1+2j)^2})  \\
\label{tilz2}
\widetilde{\cal Z}_2 ({\sf p}) &=& \frac12 \prod_{j=-\infty}^\infty 
    (1+{\sf p}^{4j^2})  \\
\label{tilz3}
\widetilde{\cal Z}_3 ({\sf p}) &=& \frac{1}{2}\prod_{j=-\infty}^\infty
        (1-{\sf p}^{(1+2j)^2})  .
\eea
Derivation of Eq.\ (\ref{zf4}) is sketched in Appendix
\ref{sec:partition2}.
The corresponding partition function $\widetilde{\cal Z}_E$
 for the asymmetric
 XXZ chain with $M$ sites at temperature $T$ can be obtained from
the $N$ even
result of Eq.\ (\ref{zf4}) with ${\sf p}$ replaced by 
${\sf p}' = \exp (-e_4 \pi^2 /(2 e_2^2 M^2
T) ) $ ; 
\be \label{zetilde}
\widetilde{\cal Z}_E =  \widetilde{\cal Z}_1 ({\sf p}') \pm \widetilde{\cal
Z}_3 ({\sf p}')
  + \widetilde{\cal Z}_2 ({\sf p}') 
\ee
with the $+$ ($-$) sign for $M$ even (odd).

Equating Eq.\ (\ref{zf2}) and (\ref{zf4}) for the 4 cases of parity of
$M$ and  $N$, we find 
\be \label{zi}
  {\cal Z}_i ( {\sf q}) = \widetilde{\cal Z}_i ( {\sf p})
\ee
for all $i$=1, 2, 3, and for any $\tau$ where ${\sf q}$ and ${\sf p}$
are related to $\tau$ by Eq.\ (\ref{sfq}) and (\ref{sfp}), respectively. 
This is an analog of the conjugate modulus  transformation of
the elliptic functions. These identities are obtained 
by calculating the physically same quantity in two independent
ways and are  confirmed
numerically. However we are not able to prove them directly. 
A byproduct of the identities is the analytic expression of $c_1$ 
mentioned below Eq.\ (\ref{c2}). 
Taking the $\tau \rightarrow \infty$ limit 
in $\ln \widetilde{\cal Z}_1 ( \exp (- \pi /\tau))$,  and evaluating
the leading order contributions using Eq.\ (\ref{sumfor}), one easily
obtains 
\be \label{limit}
\ln \widetilde{\cal Z}_1 ( \exp (- \pi /\tau)) \longrightarrow \frac12
   (\sqrt{2}-1 )\zeta(3/2) \sqrt{\tau/2}.
\ee
Comparing this with ${\cal Z}_1 (\exp(-\pi\sqrt{i\tau}))$, one obtains
the alternate result for $c_1$ as
\be \label{c12}
c_1 = (\sqrt{2}-1) \zeta(3/2)/ (4\pi).
\ee
Similar steps using $\widetilde{\cal Z}_3 (
\exp (- \pi /\tau))$ confirm the leading order behavior of 
${\cal Z}_3 ({\sf q})$ for small ${\sf q}$.

\section{discussions}
 \label{sec:discuss}
In this paper, we have presented Bethe ansatz solutions for the finite-size
corrections of the energy gaps at the antiferromagnetic phase transition
for both the asymmetric XXZ chain and the isotropic 6-vertex model or
the F model.
Furthermore, all the low-lying levels are summed to obtain explicit
expressions for partition functions. As a byproduct, interesting identities
between several infinite products are found.
Main results of this paper are Eqs. (\ref{dele}) and (\ref{delf}) for energy
gaps at $V=0$ phase boundary, Eqs. (\ref{dele2}) and (\ref{delf2}) for the
same at $H=0$ phase boundary, Eqs. (\ref{zf2}) and (\ref{zf4}) for 
the partition functions and the identities Eq.\ (\ref{zi}).

The finite-size corrections of the energy gaps are calculated using the
method developed in I (\cite{kim95}). Here, the phase function appearing
in the Bethe ansatz solution is assumed to take certain form and is
determined self-consistently. 
At $H=H_c$ and $V=0$, 
the finite-size
corrections take the form of power series in $N^{-1/2}$ 
because of the fact that the 
density of roots for the ground state  
vanishes at the end points of the root contour. The same is true
on the stochastic line treated in I but,
in the latter case, the leading
$O(N^{-1/2})$ term is absent resulting in the 
$N^{-3/2}$ scaling of energy gaps.
For $|H|<H_c$ and $V=V_c$,  the finite-size corrections come
out as a power
series in $N^{-1}$. The leading $O(N^{-1})$ term is imaginary and the real
contribution appears from $O(N^{-2})$. 
 This $N^{-2}$ scaling is the characteristic of the PT
transition and is more generic. 
The points $H=\pm H_c$ and $V=0$ of the phase boundary are special in that the
direction of the `time' of the Hamiltonian Eq.\ (\ref{ham}) and
the row-to-row transfer matrix of the lattice model 
is orthogonal to the direction of the field
and tangential to the facet boundary in the $H$-$V$ plane.

The partition function at $H=0$ and $V=V_c$ given in Eq.\ (\ref{zf4})
is actually that of an one-dimensional 
 non-relativistic free fermions with dispersion
relation $\varepsilon =A k^2$, $A$ being a constant.
 To make exact correspondence with Eq.\ (\ref{zf4}), we simply
require the momenta $k$ be even or odd integer multiple of
$\pi/L$, $L$ being the length of the system,  depending on the parity
of the total number of fermions. Thus the effective theory on 
 the $H=0$ antiferromagnetic phase boundary is that of non-relativistic
free fermions with the inverse temperature or the number of rows playing
the role of imaginary time. This is also true at $q=0$ or $q=1$ (the
dashed lines in Fig.\ 1) where
the transition can be viewed as the PT-type commensurate-incommensurate
transition \cite{dennijs88,nk1}. For such effective theory, 
the single particle wave function is $\exp
(ikx-\varepsilon t)$ where $t$ is the imaginary time. Now we consider
the effect of rotation in $x$-$t$ plane. Such transformation is 
not covered in the Schr\"{o}dinger group \cite{henkel2}
but makes sense in the context of
lattice models.  When the coordinates are rotated  
by an angle $\theta$, in order for the wave function to remain
invariant, the dispersion relation of the rotated system must change to
$\varepsilon' = i k' \tan \theta  +A (\cos \theta)^{-3}
  k'^2 +O(k'^3)$ when
$\theta\neq \pm \pi /2$ but to $\varepsilon' = \sqrt{\pm ik'/A}$
 when $\theta=\pm \pi/2$.  Since the lattice
model at $H=H_c$ and $V=0$ is obtained
by rotating that at $H=0$ and $V=V_c$ by $\pi/2$, one naturally has 
excitations with $\sqrt{ik'}$ dispersion at the $V=0$ phase boundary.
 From this, we can interprete the
$N^{-1/2}$ scaling at $H=H_c$ and $V=0$ as the feature 
arising from viewing the
ordinary PT transition with space-time coordinate
interchanged.  However, the anomalies $c_1$ and $c_2$ in the rotated
system are not explained by such simple effective theory.

\acknowledgements
The author thanks V. Rittenberg, G. v. Gehlen and  G. Albertini 
for stimulating and helpful discussions and hospitality during his visit to 
Bonn University where this work was initiated.  
He also thanks  J.D. Noh, V. Popkov and H. Lee for discussions and
comments.
This work is supported by Korea Science and Engineering Foundation through
the Center for Theoretical Physics,  Seoul National University, 
by Ministry of Education grant BSRI 96-2420 and by SNU Daewoo Research Fund.

\appendix

\section{Resursion relation of $J_\pm$}
 \label{sec:recursion}

To show the recursion relation Eq.\ (\ref{jy2}), we introduce  an
auxiliary function defined for $\Re y >0$ as
\be \label{jyy}
\tilde{J}(y)= \frac{1}{2i} \int_0^\infty \frac{\sqrt{y+it}-\sqrt{y-it}}{e^{\pi
t}-1} dt-\frac12 y^{1/2} + \frac13 y^{3/2} . 
\ee
$\tilde{J}(y)$ is related to $J_\pm (y)$ by
$J_+(y)= \sqrt{-i} \tilde{J} (-iy) $ for $\Im y >0$ and $J_-(y)=
\sqrt{i} \tilde{J}(iy)$ for  $\Im y < 0$. Eq.\ (\ref{jy2}) follows
from
the recursion relation for $\tilde{J}(y)$ which is
\be \label{jtil2}
\tilde{J}(y+2) = \tilde{J}(y) +\sqrt{y} 
\ee
for $\Re y >0$. To show Eq.\ (\ref{jtil2}), we employ 
the general sum formula \cite{dieudo}
\be \label{sumfor}
\sum_{j=0}^n f(j) = \int_0^n f(x)dx + \frac{f(n)+f(0)}{2} 
 +2\int_0^\infty \frac{\tilde{f}(n,t)-\tilde{f}(0,t)}{e^{2\pi t}-1} dt
\ee
where $\tilde{f}(x,t) = (f(x+it)-f(x-it))/2i$. This formula
 is valid if $f(x)$ is  analytic and satisfies 
$\lim_{t \rightarrow \pm \infty} e^{-2\pi |t|}f(x+it) =0$ 
in the strip $-\delta <\Re x < n+\delta$ for some $\delta
>0$.
Applying Eq.\ (\ref{sumfor}) to a trivial sum $\sum_{j=0}^1
\sqrt{y+2j}$, one then obtains Eq.\ (\ref{jtil2}).

\section{Partition function at $H=H_c$ and $V=0$}
\label{sec:partition}

In this appendix, we evaluate the partition function of the lattice model
at the $V=0$ phase boundary.
That of the asymmetric XXZ chain follows simply redefining the nome
and using the result of the lattice model for $M$ even. 
For simplicity, we consider the case of $M$ even and mention the other case 
at the end. Since $M$ is even, from Eq.\ (\ref{zf}), we are to evaluate 
\be \label{zff}
{\cal Z}_F = \sum_{(r,m,{\cal P})} \exp ( -\pi \sqrt{\tau} {\cal
Y}_1^{(r,m,{\cal P})} ),
\ee
where $\tau$ is given in Eq.\ (\ref{tau}). 

First consider the case $(2r+m)$ even. Further, if $m$ is in the range
$-(2r+1)< m \leq (2r+1)$,  $ {\cal Y}_1^{(r,m,0)}$ is
 given in Eq.\ (\ref{rm0e})  and we have
\be \label{front}
\exp ( -\pi \sqrt{\tau} {\cal
Y}_1^{(r,m,0)} ) = ( {\sf q}\bar{\sf q})^{-c_1}
\prod_{j=0}^{r-m/2-1} \bar{\sf q}^{\sqrt{1+2j}} 
\prod_{j=0}^{r+m/2-1} {\sf q}^{\sqrt{1+2j}}
\ee
where ${\sf q}= \exp (-\pi \sqrt{i\tau})$ and $\bar{\sf q}= \exp (-\pi
\sqrt{-i\tau})$. From Eq.\ (\ref{caly}), one
notes that $\exp ( -\pi \sqrt{\tau} {\cal Y}_1^{(r,m,{\cal P})} )$ is 
 a product of $\exp ( -\pi \sqrt{\tau} {\cal
Y}_1^{(r,m,0)})$ and factors coming
 from particle and hole excitations. A particle at position $p_k$ with
$2p_k < (2r-m+1)$ contributes a factor
 $\bar{\sf q}^{-\sqrt{1+2j}}$ where $j=
r-m/2-p_k=0,1,\ldots,(r-m/2-1)$, while that with $2p_k \geq (2r-m+1)$
a factor ${\sf q}^{\sqrt{1+2j}}$ where $j=p_k-r+m/2-1=0,1,
\ldots,\infty$. A hole at $h_k$ gives $\bar{\sf q}^{\sqrt{1+2j}}$ where
$j=h_k+r-m/2-1=r-m/2,r-m/2+1,\ldots,\infty$. Summing over all particle-hole
configurations at the right-end of the Fermi sea corresponds to evaluating
a free fermion grand-canonical partition function under the constraint that
the total number of particles should be the same as that of holes.
If we denote the fugacity of each particle (hole) by $z$ ($z^{-1}$), the
desired sum is obtained by projecting out the $z^0$ term of the free
fermion grand-canonical partition function which is 
\be \label{xi1}
\Xi_1 (z)= \prod_{j=0}^{r-m/2-1}(1+z \bar{\sf q}^{-\sqrt{1+2j}})
 \prod_{j=0}^{\infty}(1+z {\sf q}^{\sqrt{1+2j}})
\prod_{j=r-m/2}^\infty (1+z^{-1} \bar{\sf q}^{\sqrt{1+2j}}).
\ee
$\Xi_n (z)$ of this Appendix and Appendix \ref{sec:partition2} are not to
be confused with $\Xi (b)$ defined in Eq.\ (\ref{xib}).
We introduce a function $A(z)$ by
\be \label{az}
A(z)= \prod_{j=0}^{\infty}(1+z {\sf q}^{\sqrt{1+2j}})
\prod_{j=0}^\infty (1+z^{-1} \bar{\sf q}^{\sqrt{1+2j}}),
\ee
and denote its Fourier coefficients by $A_n$; $A(z)=\sum_n A_n z^n$. Then
multipying and dividing by a factor $\prod_{j=0}^{r-m/2-1} (1+ z^{-1}
\bar{\sf q}^{\sqrt{1+2j}})$ on the right-hand side of Eq.\ (\ref{xi1}), we
can put $\Xi_1$ in the form
\be \label{xi12}
\Xi_1(z)=  
  z^{r-m/2} A(z)\prod_{j=0}^{r-m/2-1} \bar{\sf q}^{-\sqrt{1+2j}} .
\ee
Thus the coefficient of $z^0$ of $\Xi_1(z)$ is $A_{-r+m/2}
 \prod_{j=0}^{r-m/2-1}
\bar{\sf q}^{-\sqrt{1+2j}} $. Note that the product over $j$ in Eq.\
(\ref{xi12}) is cancelled by that in Eq.\ (\ref{front}).
 Summation over particle-hole 
configurations at the left-end of the Fermi sea proceeds in exactly the
same way with $m$ replaced by $-m$ and ${\sf q}$ by $\bar{\sf q}$.
Thus, the partition function after
summation over all excitations from each $m$-shifted state is given as
\be \label{pf2}
\sum_{\cal P} \exp ( -\pi \sqrt{\tau} {\cal Y}_1^{(r,m,{\cal P})} )
 = ( {\sf q} \bar{\sf q})^{-c_1} A_{-r+m/2}\bar{A}_{-r-m/2}
\ee 
for $-(2r+1) < m \leq (2r+1)$ where $\bar{A}_n$ stands for the complex
conjugate of $A_n$. Similar calculations for each case of 
$m > (2r+1)$ and $m \leq -(2r+1)$ show that Eq.\ (\ref{pf2}) holds for all
$m$, provided  $(2r+m)$ is even.  We then sum  the right-hand side of 
Eq.\ (\ref{pf2}) over all  even (odd) $m$'s for $2r$ even (odd). However, we
note that $\sum_m A_{-r+m/2} \bar{A}_{-r-m/2}$ is simply the coefficient of
$z^{-2r}$ in $|A(z)|^2$. To obtain the partition function in the $(2r+m)$ even
sector, we finally 
 sum over all integer (half integer) values of
$r$ if $N$ is even (odd). Even though above derivations assumed $r\geq0$,
we may sum over negative $r$'s by the symmetry, $|A(z)|^2=|A(z^{-1})|^2$.
 Since the sum over $r$  is equivalent to summing 
over all coefficients of even (odd) powers
of $z$ in $|A(z)|^2$, it can be written as $(A(1)^2+A(-1)^2)/2$
($(A(1)^2-A(-1)^2)/2$) for $N$ even (odd).
  Noting that $({\sf q}\bar{\sf q})^{-c_1} 
A(1)^2/2= {\cal Z}_1 ({\sf q})$ and
$ ({\sf q}\bar{\sf q})^{-c_1} A(-1)^2/2= {\cal Z}_2 ({\sf q})$,
we have the partition function in the $(2r+m)$ even
sector as ${\cal Z}_1 ({\sf q})\pm  {\cal Z}_2 ({\sf q})$ with $+$ ($-$)
sign for $N$ even (odd).

Next consider the case of $(2r+m)$ odd. If $m$ is in the range
 $-(2r+1)< m \leq (2r+1)$,  $ {\cal Y}_1^{(r,m,0)}$
 is given in Eq.\ (\ref{rm0o})  and we have
\be \label{front2}
\exp ( -\pi \sqrt{\tau} {\cal Y}_1^{(r,m,0)} ) = ( {\sf q}\bar{\sf q})^{c_2}
\prod_{j=0}^{r-(m+1)/2} \bar{\sf q}^{\sqrt{2j}} 
\prod_{j=0}^{r+(m-1)/2} {\sf q}^{\sqrt{2j}}
\ee
 A particle at position $p_k$ with
$2p_k \leq (2r-m+1)$ contributes a factor
 $\bar{\sf q}^{-\sqrt{2j}}$ where $j=
r-(m-1)/2-p_k=0,1,\ldots,r-(m+1)/2$, while that with $2p_k > (2r-m+1)$
a factor ${\sf q}^{\sqrt{2j}}$ where $j=p_k-r+(m-1)/2=1,2,
\ldots,\infty$. A hole at $h_k$ gives $\bar{\sf q}^{\sqrt{2j}}$ where
$j=h_k+r-(m+1)/2=r-(m-1)/2,r-(m-1)/2+1,\ldots\infty$. 
Thus the free
fermion grand-canonical partition function for excitations at the
right-hand side  is 
\be \label{xi2}
\Xi_2 (z)= \prod_{j=0}^{r-(m+1)/2}(1+z \bar{\sf q}^{-\sqrt{2j}})
 \prod_{j=1}^{\infty}(1+z {\sf q}^{\sqrt{2j}})
\prod_{j=r-(m-1)/2}^\infty (1+z^{-1} \bar{\sf q}^{\sqrt{2j}}).
\ee
In line with the $(2r+m)$ even sector, we introduce a function $B(z)$ by
\be \label{bz}
B(z)= \prod_{j=1}^{\infty}(1+z {\sf q}^{\sqrt{2j}})
\prod_{j=0}^\infty (1+z^{-1} \bar{\sf q}^{\sqrt{2j}}),
\ee
and denote its Fourier coficients by $B_n$; $B(z)=\sum_n B_n z^n$. Then
$\Xi_2$ becomes
\be \label{xi22}
\Xi_2(z)=  
  z^{r-(m-1)/2} B(z) \prod_{j=0}^{r-(m+1)/2} \bar{\sf q}^{-\sqrt{2j}} 
\ee
and one sees that 
contribution from particle-hole excitations at the right-end is proportional to
$B_{-r+(m-1)/2}$. Adding that of the left-end, we have  
the intermediate partition function as 
\be \label{pf3}
\sum_{\cal P} \exp ( -\pi \sqrt{\tau} {\cal Y}_1^{(r,m,{\cal P})} )
 = ( {\sf q} \bar{\sf q})^{c_2} B_{-r+(m-1)/2} \bar{B}_{-r-(m+1)/2}
\ee 
for $-(2r+1) < m \leq (2r+1)$ and $(2r+m)$ odd.
Similar calculations for each case of 
$m > (2r+1)$ and $m \leq -(2r+1)$ again show
 that Eq.\ (\ref{pf3}) holds for all
$m$, provided  $(2r+m)$ is odd.  Proceeding as above, we sum over
$m$ and $r$ in Eq.\ (\ref{pf3}) using the symmetry $z|B(z)|^2=z^{-1}
|B(z^{-1})|^2$  
and finally find the partition function in $(2r+m)$ odd sector
as $( {\sf q} \bar{\sf q})^{c_2}(B(1)^2+B(-1)^2)/2$ for $2r$ odd and 
 $( {\sf q} \bar{\sf q})^{c_2}(B(1)^2-B(-1)^2)/2$ for $2r$
even. But $B(-1)=0$ so that it becomes $( {\sf q} \bar{\sf q})^{c_2}
B(1)^2/2={\cal Z}_3({\sf q})$ for
both parity of $N$. Adding
this to that of $(2r+m)$ even sector, we have Eq.\ (\ref{zf2}) for $M$ even.

When $M$ is odd, we need to insert a factor $(-1)^m$  to Eqs.\ (\ref{pf2})
and (\ref{pf3}). This does not complicate the sums since the sums over $m$
in each sector are in steps of 2 and the sign factor is equivalent to
$(-1)^{2r}$ for the case of Eq.\ (\ref{pf2}) and $(-1)^{2r+1}$ for 
the case of Eq.\ (\ref{pf3}). These extra sign factors then give 
the full partition function as given in 
Eq.\ (\ref{zf2}) for $M$ odd.

\section{Spectra at $H=0$ and $V=V_c$}
\label{sec:spectra3}

In this section, we derive $\Delta E$ and $\Delta F$ 
at the phase boundary of $H=0$.
Although our main interest
is for $H=0$, discussions up to Eqs.\
 (\ref{dele4}) and (\ref{delf4}) hold for general $0 \leq H < H_c$.

When $|H| < H_c$, $Z_\infty$ given by Eq.\ (\ref{zinfty}) has non-vanishing
first derivative at $x_0 \equiv \exp ( \pi i +\lambda - b)$,  the
end point of the root distribution in bulk limit, which is
 related to $H$ by Eq.\ (\ref{h}).  Thus our starting
ansatz for $Z_N^{-1}(x)$ is, instead of Eq.\ (\ref{zninv}), 
\be \label{zninv2}
Z_N^{-1}(\pm \pi q + \pi \xi /N) = x_0^\pm + a_2 \pi(y_\pm - i \xi)/N
  + a_4 \pi^2 (y_\pm - i\xi)^2/N^2 + \cdots
\ee
where $x_0^+=x_0$ and $x_0^-=x_0e^{-2\pi i}$.
Here, $y_\pm$, $a_2$ and $a_4$ are to be determined
self-consistently. We then follow the same line
of steps as in  Section \ref{sec:energy}. However, 
as we have seen in Section \ref{sec:crossover}, 
the levels not satisfying the condition Eq.\ (\ref{cond})
become higher in energy by $O(1)$ than those which do. This is
because the roots of the Bethe ansatz equation, $x_j$ or $x_{Q-j}$,
 for $j$ finite,  may not remain close to $x_0^\pm$. So,
  Eq.\ (\ref{zninv2}) is useful  only to the levels which do satisfy 
Eq.\ (\ref{cond}). 
 Repeating the algebra, the sum Eq.\
(\ref{sf}) is then found to take the form 
\bea \label{sf3}
S[f] &=& \frac{N}{2\pi i} \oint f(x) R_N(x) \frac{dx}{x} + 
              (m+iy_+/2)f(x_0^+)-(m+iy_-/2) f(x_0^-) \nonumber \\
  &\  &+ f'(x_0) \Phi + f''(x_0) \Psi  +O(N^{-3})
\eea
where $\Phi$ and $\Psi$ are short notations for
\bea \label{cd}
 \Phi &=& a_2 \pi Y_2(y_+, y_-) /N + a_4\pi^2 Y_4(y_+,y_-)/ N^2 , \\
 \Psi &=&  \frac12 a_2^2 \pi^2 Y_4 (y_+, y_-)/N^2
\eea
with
\bea \label{y2n}
Y_{2n} (y_+,y_-) &=& -i J_+^{(n)} (y_+ + i-2mi) + i J_-^{(n)} (y_--i-2mi) 
 \nonumber \\
 &\ & + \sum_{k=1}^{n_+} \left\{ (y_+ +i(1-2m-2p_k))^n - (y_+-i(1+2m-2h_k))^n
  \right\} \nonumber \\
 &\ & + \sum_{k=1}^{n_-} \left\{ (y_--i(1+2m-2\bar{p}_k))^n -
  (y_- +i(1-2m-2\bar{h}_k))^n \right\} .
\eea
Here, $J_\pm^{(n)} (y)$ are a generalization of Eq.\ (\ref{jy}):
\be \label{jny}
J_\pm^{(n)} (y) = \frac12 \int_0^\infty  \frac{(y+t)^n-(y-t)^n}{e^{\pi t}
-1} d t \pm \frac{i}{2} y^n - \frac{1}{2(n+1)} y^{n+1} .
\ee
When $y_+=y_-$ and $m=0$, $(-1)^nY_{2n}(y,y)$ reduces to $Y_{2n}(y)$ of I.

Using this general sum formula, we obtain the  solution for $R_N$ as 
\be \label{rn3}
R_N (x) = R_\infty (x)+ \frac{r}{N}+ \frac{1}{N} \sum_{n\neq0} 
\frac{\alpha^{|n|}}{1+\alpha^{|n|}} \left( \frac{x}{x_0}\right)^n
 \left( \frac{y_+-y_-}{2i} + \frac{n }{x_0} \Phi-\frac{n(n+1)}{x_0^2} \Psi
 \right) + O(N^{-4}),
\ee
and the self-consistency equations for $y_\pm$, $a_2$ and $a_4$ as
\bea \label{yeq2}
\pm i \pi q  + y_\pm \frac{\pi}{N} &=& \pm i (1-q) \pi +
   2H-\lambda+ 2 \sum_{n\neq 0}
 \frac{\alpha^n}{1+\alpha^n} \frac{x_0^n}{n} \nonumber \\
&\ & + (1+ \frac{2r}{N}-\frac{y_+-y_-}{2Ni}) \ln |x_0|
  + (2im-\frac{y_++y_-}{2}) \frac{\pi}{N}  \nonumber \\
&\ & + 2 e_0x_0^{-1} (\Phi-\Psi/x_0)N^{-1}  
   + O(N^{-4}), 
\eea
\bea \label{a24}
 -a_2^{-1}&=&  iZ_\infty'(x_0) + 2re_0x_0^{-1} N^{-1}
  +O(N^{-3}), \\
 2a_4 a_2^{-3} &=& iZ_\infty''(x_0) + O(N^{-1})
\eea
where $Z_\infty(x)$ is given in Eq.\ (\ref{zinfty}) and $y_+-y_-=4ir$ which 
follows from Eq.\ (\ref{yeq2}) has been used in Eq.\ (\ref{a24}).
 From Eq.\ (\ref{yeq2}),
one gets, after using Eq.\ (\ref{2h}),  
\be \label{ypm2}
y_\pm = (\pm 2r +m)i + \pi^{-1} e_0 (\Phi/x_0-\Psi/x_0^2) + O(N^{-3}).
\ee
Applying Eq.\ (\ref{sf3}) with Eqs.\ (\ref{rn3}) and (\ref{ypm2}) to 
Eqs. (\ref{e}) and (\ref{loglam}), we then obtain after some algebra, 
\be  \label{dele4}
\Delta E = 2r \Xi '(b) + \Xi '' (b)
  (x_0^{-1}\Phi-x_0^{-2} \Psi )-\Xi '''(b) x_0^{-2} \Psi  +O(N^{-3})
\ee
for the chain and
\be \label{delf4}
\Delta F = -\pi i m -2r (V-\Xi (\lambda -b)) 
  - \Xi '(\lambda -b) (x_0^{-1}\Phi-x_0^{-2}\Psi)
  -\Xi '' (\lambda -b) x_0^{-2} \Psi + O(N^{-3})
\ee
for the lattice model where $x_0 = - \exp(\lambda -b)$ and 
$\Xi(b)$ is  defined in Eq.\ (\ref{xib}).
The $O(1)$ terms in Eqs.\ (\ref{dele4}) and (\ref{delf4}) imply that
the critical field $V_c$ as a function of $H$ 
is given by $\Xi '(b)$ for the chain and by $\Xi (\lambda -b)$ for
the lattice. 
 These $O(1)$ terms
are  offset
by the vertical field at the phase boundary. Eqs.\ (\ref{dele4}) and
(\ref{delf4}) contain, through $\Phi$, an $O(N^{-1})$ term  proportional to 
$Y_2(y_\pm = \pm 2r i +mi)$ which is purely imaginary. Thus the real parts
of the energy gaps at the antiferromagnetic phase boundary 
are $O(N^{-2})$.

Up to here  $H$ is general. For simplicity, we now specialize to $H=0$ at
corresponding values of the  critical vertical field. When $H=0$, $x_0=
-e^\lambda$, $b=0$  and $a_2= e^{\lambda}/e_2+O(N^{-1})$.
Setting $b=0$ in Eqs.\ (\ref{dele4}) and (\ref{delf4}),
one notes that  
the imaginary $O(N^{-1})$ terms all vanish simplifying the
results considerably. Final expression for $\Delta E$ and $\Delta F$ then
becomes as given in Eqs.\ (\ref{dele2}) and (\ref{delf2}), respectively,
where 
${\cal Y}_4^{(r,m,{\cal P})}$ is the value of $Y_4 (y_+,y_-)$ evaluated
at the zero-th order solution for $y_\pm$, i.e. $Y_4(y_\pm=i(\pm 2r + m))$.

\section{Partition functions at $H=0$ and $V=V_c$}
\label{sec:partition2}
 In this Appendix, we derive Eq.\ (\ref{zf4}).
Consider the lattice model on $M\times N$ lattice with $M$ and $N$ even.
Then from Eqs.\ (\ref{zf3}) and (\ref{sfp}), we have
\be \label{zf6}
{\cal Z}_F = \sum_{(r,m,{\cal P})} {\sf p}^{{\cal Y}_4^{(r,m,{\cal P})}} 
\ee
with ${\cal Y}_4^{(r,m,{\cal P})}$ given in Eq.\ (\ref{caly4}) and
the sum is for $r=0,1,\ldots$ 
under the restriction Eq.\ (\ref{cond}). Considering first $(2r+m)$ even
sector, from Eq.\ (\ref{caly4}), one sees
that a particle (hole) at position $p_k$ ($h_k$) contributes ${\sf
p}^{-(1+2j)^2}$ (${\sf p}^{(1+2j)^2}$) with $0\leq j\leq r-m/2-1$ (
$j \geq r-m/2$). Thus the free fermion grand-canonical
 partition function for the
right-hand side excitation is 
\bea \label{xi3}
\Xi_3 (z)&=& \prod_{j=0}^{r-m/2-1}(1+z {\sf p}^{-(1+2j)^2} )
           \prod_{j=r-m/2}^\infty (1+z^{-1} {\sf p}^{(1+2j)^2} )
\nonumber \\
  &=& z^{r-m/2} C(z)\prod_{j=0}^{r-m/2-1} {\sf p}^{-(1+2j)^2}
\eea
where
\be \label{c}
C(z) = \prod_{j=0}^\infty (1+z^{-1} {\sf p}^{(1+2j)^2} )=\sum_{n=0}^\infty
C_n z^{-n} .
\ee
Thus the coefficient of $z^0$ in $\Xi_3(z)$ can be written as $C_{r-m/2} 
\prod_{j=0}^{r-m/2-1} {\sf p}^{-(1+2j)^2}$. 
The left-hand side excitations give the same result with $m\rightarrow -m$.
Using the fact
that $\sum_{j=0}^{r-m/2-1}(1+2j)^2 + \sum_{j=0}^{r+m/2-1}(1+2j)^2
 = 8r^3/3-2r/3+2rm^2$, we note that the product in the second line of
Eq.\ (\ref{xi3})  
is cancelled by that coming from the first terms of
${\cal Y}_4^{(r,m,{\cal P})}$. Thus we have 
\be \label{pf4}
 \sum_{\cal P}  {\sf p}^{{\cal Y}_4^{(r,m,{\cal P})}} = C_{r-m/2}C_{r+m/2}.
\ee
Summing Eq.\ (\ref{pf4}) over $m$ from $-2r$ to $+2r$ in steps of 2 and
over $r\geq0$, we find the partition function in the $(2r+m)$ even sector as
$(C(1)^2+C(-1)^2)/2=\widetilde{\cal Z}_1 ({\sf p}) + \widetilde{\cal Z}_3
({\sf p})$.

Next consider the $(2r+m)$ odd sector. In this case the free fermion 
grand-canonical
partition function for the right-hand side excitations is 
\be \label{xi4}
 \Xi_4 (z) = \prod_{j=0}^{r-(m+1)/2} (1+z {\sf p}^{-4j^2} )
           \prod_{j=r-(m-1)/2}^\infty (1+z^{-1} {\sf p}^{4j^2} ) 
\ee
while that for the left-hand side is
\be \label{xi5}
 \widetilde{\Xi}_4 (z) = \prod_{j=1}^{r+(m-1)/2} (1+z {\sf p}^{-4j^2} )
           \prod_{j=r+(m+1)/2}^\infty (1+z^{-1} {\sf p}^{4j^2} ).
\ee
Note that the $j=0$ term is excluded in Eq.\ (\ref{xi5}) due to the
strict inequality in Eq.\ (\ref{cond}). 
Defining 
\bea \label{dd}
D(z) &=& \prod_{j=0}^\infty (1+z^{-1} {\sf p}^{4j^2} )=\sum_{n=0}^\infty
D_n z^{-n} \\
 \widetilde{D} (z) &=& \prod_{j=1}^\infty
 (1+z^{-1} {\sf p}^{4j^2})=\sum_{n=0}^\infty
\widetilde{D}_n  z^{-n},
\eea
we can write 
\be \label{pfr54}
 \sum_{\cal P}  {\sf p}^{{\cal Y}_4^{(r,m,{\cal P})}} = D_{r-(m-1)/2}
\widetilde{D}_{r+(m-1)/2}
\ee
where $\sum_{j=0}^{r-m/2-1/2}(2j)^2 + \sum_{j=1}^{r+m/2-1/2}(2j)^2
 = 8r^3/3-2r/3+2rm^2$
has been used. We then sum Eq.\ (\ref{pfr54})
 over $m$  from $-2r+1$ to $2r+1$ in steps of 2
and finally over $r \geq 0$ to obtain the remaining piece of the partition
function as $(D(1)\widetilde{D}(1)+D(-1)\widetilde{D}(-1))/2=
\widetilde{\cal Z}_2 ({\sf p})$. We thus arrive at Eq.\ (\ref{zf4}) for $M$
and $N$ even. For other parities of $M$ and $N$, we  repeat 
above steps with only minor changes and find  Eq.\ (\ref{zf4}).

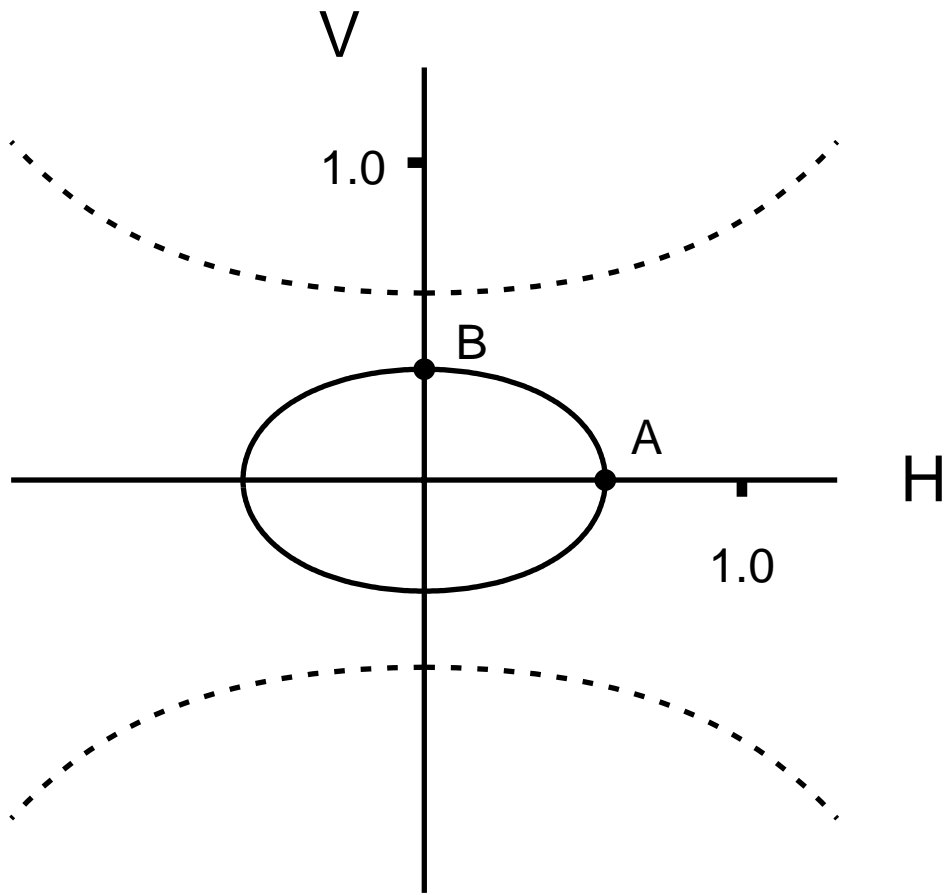
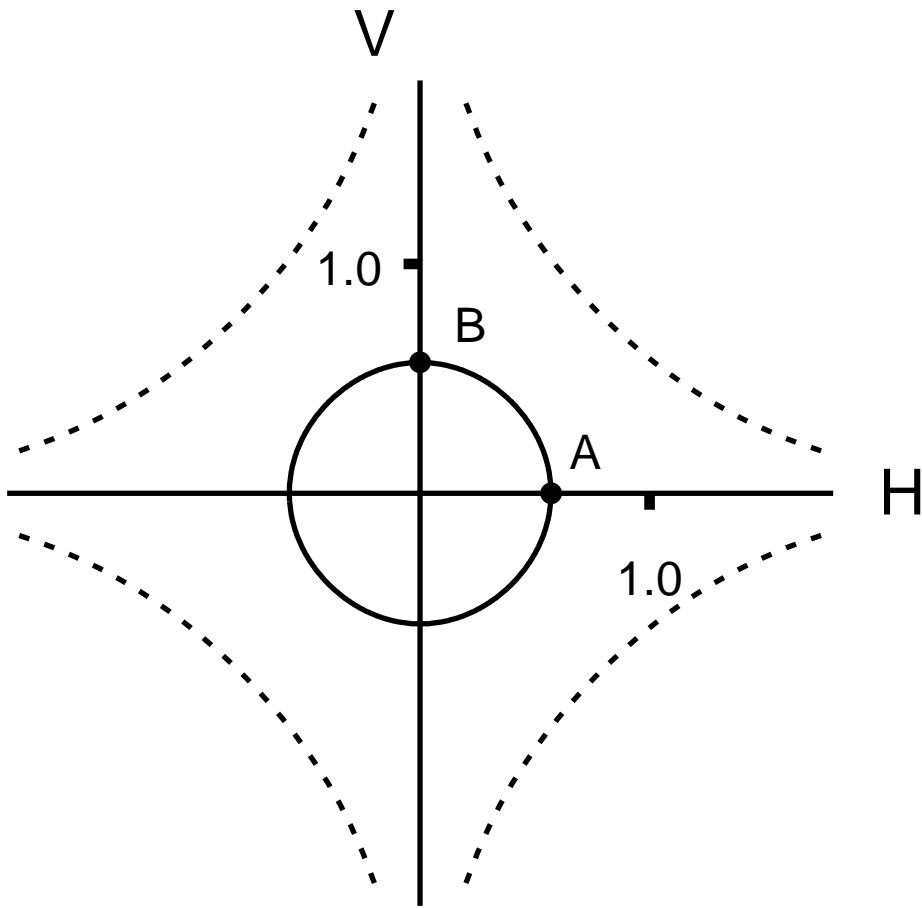
\begin{figure}
\caption{(a) Phase diagram in fields of the asymmetric XXZ chain for
$\lambda=2.5$. The solid curve is the antiferromagnetic phase boundary
and the dashed ones are the ferromagnetic phase boundary. The region
between the curves is the critical phase. The point A (B)  is the 
$H=H_c$, $V=0$ ($H=0$, $V=V_c$) phase boundary disussed in Section
\ref{sec:spectra} (\ref{sec:spectra2}).
(b) The same as in (a) for the isotropic 6-vertex model (F-model).}
\end{figure}
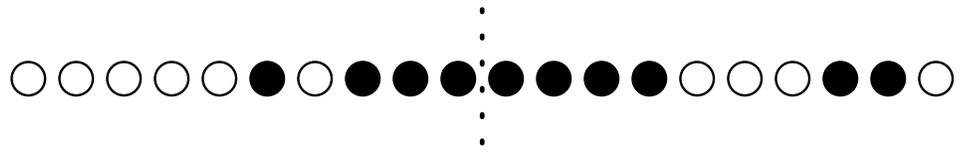
\begin{figure}
\caption{Examples of energy levels for $N=20$ and $Q=10$ ($r=0$). Filled
circles denotes integers included in $\{I_j\}$. (a) The ground state
(0,0,0) (b) The 1-shifted level (0,1,0) (c) An excited level from (b)
which has two particle-hole pairs at the right-end of the Fermi sea
(with positions 
$p_1$=2, $p_2$=3, $h_1$=1 and  $h_2$=2) and one particle-hole pair at
the left-end (with $\bar{p}_1$=1 and $\bar{h}_1$=1). }
\end{figure}

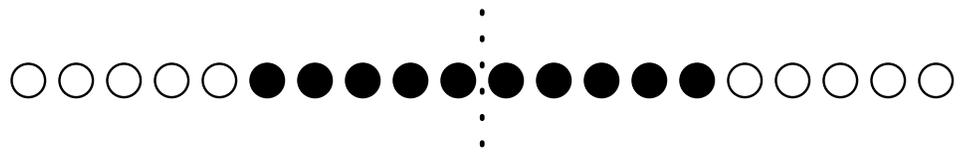
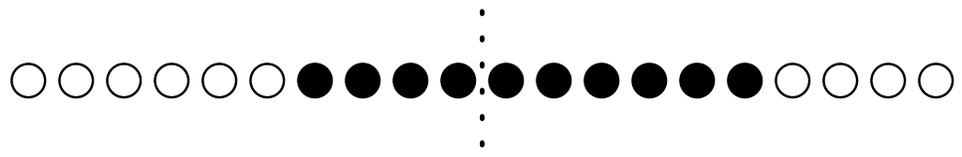
\end{document}